\documentclass[10pt]{article}
\usepackage[english]{babel}		
\usepackage{float}     
\usepackage{braket}
\usepackage{subfigure}
\usepackage{graphicx}
\usepackage{caption}
\usepackage{physics} 
\usepackage{authblk}
\usepackage{hyperref}
\usepackage{enumerate} 
\usepackage{pgfplots}
\usepackage{pgfplotstable}
\usepackage{tikz,pgfplots}
\usepackage{amsmath}  
\usepackage{wasysym} 
\date{\today}
\usepackage{geometry}
\pgfplotsset{compat=1.14}
\definecolor{darkblue}{rgb}{0.0, 0.0, 0.55} 
\definecolor{mycolor}{rgb}{0, 0, 128}  
\hypersetup{colorlinks=true,linkcolor=darkblue,filecolor=magenta,
urlcolor=darkblue,}

\begin{document}
\title{\textbf{\textcolor{darkblue}{Revisiting comparison between entanglement measures for two-qubit pure states}}}  
\author[1]{Ashutosh Singh}
\author[2,3]{Ijaz Ahamed}
\author[3]{Dipankar Home}
\author[1]{Urbasi Sinha*} 
\affil[1]{Light and Matter Physics Group, Raman Research Institute, Sadashivanagar, Bangalore 560080, India.}
\affil[2]{Indian Institute of Science Education and Research Mohali, Sector 81, Mohali, 140306, India.}
\affil[3]{Center for Astroparticle Physics and Space Science (CAPSS), Bose Institute, Kolkata 700091, India.}
\affil[*]{Corresponding author: \href{mailto: usinha@rri.res.in}{usinha@rri.res.in}}
\date{\today}
\maketitle  

\begin{abstract}
Given a non-maximally entangled state, an operationally significant question is to quantitatively assess as to what extent the state is away from the maximally entangled state, which is of importance in evaluating the efficacy of the state for its various uses as a resource. It is this question which is examined in this paper for two-qubit pure entangled states in terms of different entanglement measures like Negativity (N), Logarithmic Negativity (LN), and Entanglement of Formation (EOF). Although these entanglement measures are defined differently, to what extent they differ in quantitatively addressing the earlier mentioned question has remained uninvestigated. Theoretical estimate in this paper shows that an appropriately defined parameter characterizing the fractional deviation of any given entangled state from the maximally entangled state in terms of N is quite different from that computed in terms of EOF with their values differing up to $\sim 15\%$ for states further away from the maximally entangled state. Similarly, the values of such fractional deviation parameters estimated using the entanglement measures LN and EOF, respectively, also strikingly differ among themselves with the maximum value of this difference being around $23\%$. This analysis is complemented by illustration of these differences in terms of empirical results obtained from a suitably planned experimental study. Thus, such appreciable amount of quantitative non-equivalence between the entanglement measures in addressing the experimentally relevant question considered in the present paper highlights the requirement of an appropriate quantifier for such intent. We indicate directions of study that can be explored towards finding such a quantifier.
\end{abstract}

\textcolor{darkblue}{\section{{Introduction: Background and Motivation}}}

Entanglement lies at the core of Quantum Foundational studies leading to Information Theoretic applications and forms the bedrock of Quantum Computation. One of the key concepts used for studying entanglement is what is known as Entanglement Measure (EM) which is invoked for quantifying entanglement. For this purpose, different EMs have been proposed. It was argued by Bennett \textit{et al.} [\ref{01}, \ref{02}] that the Entanglement of Formation (EOF), intended to quantify the resources needed to create a given entangled state, satisfies the criterion of being nonincreasing under local operations and classical communication (LOCC); for bipartite pure states it is given by the von Neumann entropy of reduced density matrix relevant to either Alice or Bob, also known as Entanglement Entropy. Justification of the above EM from the thermodynamic considerations was given by Popescu \textit{et al.} [\ref{03}], followed by a comprehensive analysis due to Vedral \textit{et al.} [\ref{04},\ref{05}] and Vidal [\ref{06}] who argued that only one EM is not sufficient to completely quantify entanglement of pure states for bipartite systems. Subsequently, $\dot{Z}$yczkowski \textit{et al.} [\ref{07},\ref{08}] defined Negativity as a ``quantity capable of measuring a degree of entanglement". Later, Negativity was proved to be a valid EM [\ref{09}-{\ref{11}] by showing that it is an entanglement monotone, i.e., nonincreasing under LOCC.\\

In this paper, we have used the particular expression of Negativity (N) given by Vidal and Werner [\ref{10}], who also defined another quantity called Logarithmic Negativity (LN = $\log_{2}(2\text{N}+1)$) as a valid EM which exhibits a form of monotonicity under LOCC (non-increasing under deterministic distillation protocols) and signifies an upper bound of distillable entanglement. In a separate line of work, for bipartite qubit states, Wootters [\ref{12}] expressed EOF as a monotonic function of a quantity called `Concurrence' and argued that Concurrence can also be regarded as a measure of entanglement. Note that, for bipartite pure qubit states, Concurrence is twice of Negativity [\ref{13}], thus implying that EOF is also a monotonic function of N and LN for such states. The Fig.~\ref{fig1} below shows the comparison of different EMs for  a two-qubit pure state: $|\psi\rangle=c_0|00\rangle+c_1|11\rangle$, where $c_0$ and $c_1$ are the Schmidt coefficients. It is worth noting that (a) Concurrence and twice Negativity, and (b) Entanglement of Formation and Entanglement Entropy match with each other. Therefore, in this work, while considering essentially two-qubit pure states,  we focus on N, LN and EOF as the relevant EMs as these are the ones which do not overlap.
	  
\begin{figure} [h!] 
\begin{center}
\includegraphics[clip, trim=0cm 0cm 0cm 0cm, width=0.65 \columnwidth]{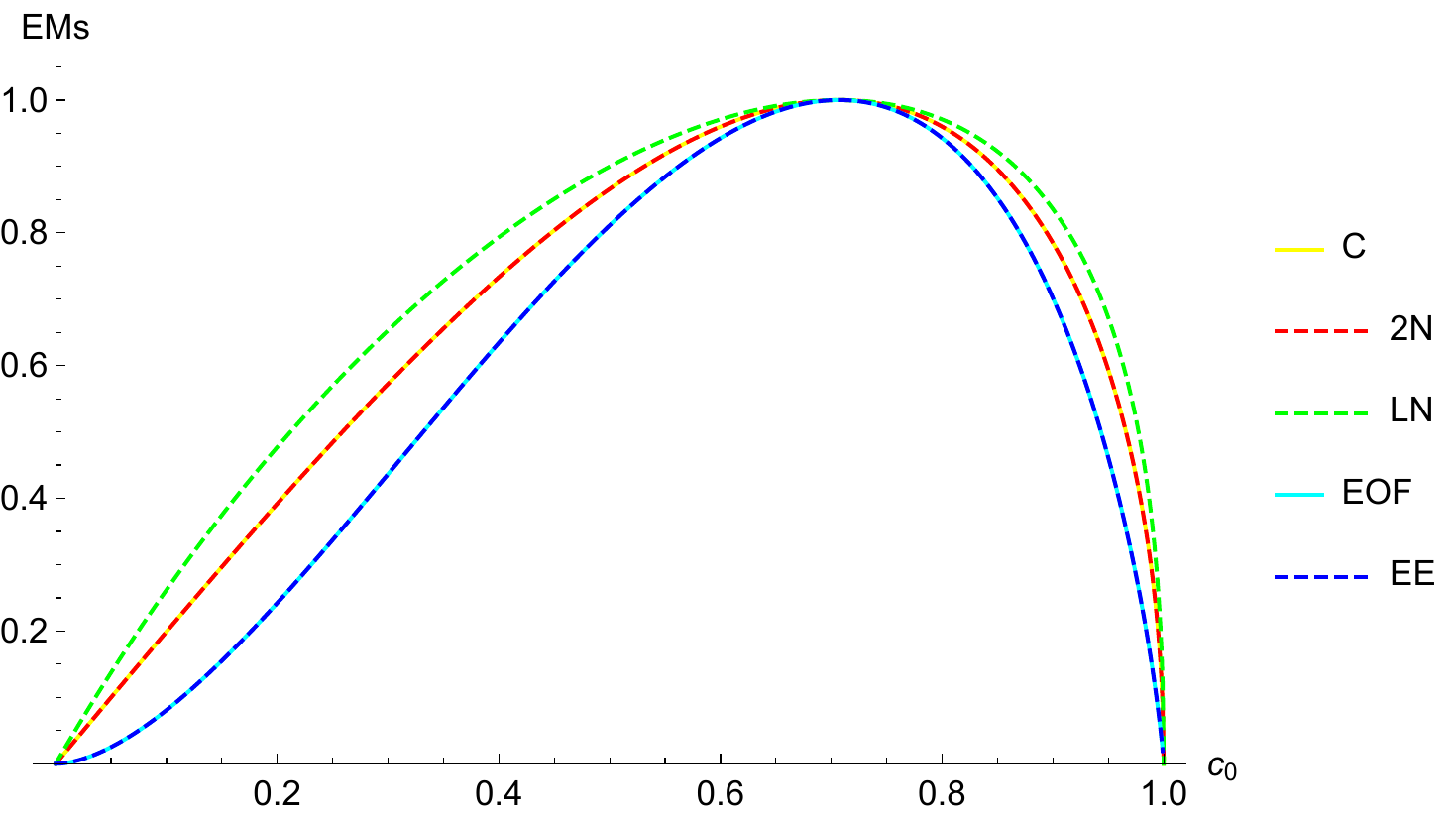}
\end{center}
\caption{\textit{A comparison of different entanglement measures with respect to the state parameter $c_0$ for two-qubit pure states. Here, C, N, LN, EOF, and EE denote Concurrence, Negativity, Logarithmic Negativity, Entanglement of Formation, and Entanglement Entropy, respectively.}}
\label{fig1}
\end{figure}
	 
From the Fig.~\ref{fig1}, it can be seen that although for any given two-qubit pure state, the values of N, LN and EOF differ among themselves, these EMs are monotonic with respect to each other. Hence, for comparing the amount of entanglement between two-qubit pure states, N, LN and EOF are all equivalent in the sense that all these EMs give the same result in answering the question as to whether a given two-qubit pure state is more (less) entangled than any other state. On the other hand, in this paper, an operationally relevant different question is addressed; i.e., of quantifying the percentage deviation of a given state from the maximally entangled state - it is in this context, we consider the issue of comparison between the various EMs. To this end, by appropriately defining the measures of such percentage deviations in terms of N, LN and EOF, respectively, we present in Section~\ref{secII} the theoretical estimates of these measures for the general class of two-qubit pure states by varying values of the Schmidt coefficients. This thorough study reveals a considerable amount of disagreement between the computed measures of percentage deviations from the maximally entangled state using N, LN and EOF, respectively. \\

A complementary line of study of this issue is then presented in Section~\ref{secIII} by considering a range of states produced in a relevant experimental study for which the quantities N, LN and EOF are determined from the density matrix reconstructed using quantum state tomography. The results obtained in this way confirm significant quantitative non-equivalence between these EMs in capturing the extent to which a given non-maximal entangled state is deviating from the maximally entangled state. This finding, therefore, underscores the need for identifying an appropriate quantifier for addressing such an empirically relevant question even in the simplest case of  $2\otimes 2$ composite systems.\\

In Section~\ref{secIV}, we extend the current analysis to higher dimensional systems; bipartite pure qutrit states and tripartite pure qubit states, in particular. In Section~\ref{secV}, we discuss possible directions of study for addressing this issue based on the various suggested ideas of `distance measure' between quantum states, as well as a line of study is outlined in terms of the deviation of the maximum value of the violation of Bell-CHSH inequality for a given state from that  corresponding to the maximally entangled state. Further, indications have been given about the way  the results obtained from such studies can be compared with those obtained from different EMs. This is followed by concluding remarks in Section~\ref{secVI}.
	 
\textcolor{darkblue}{\section{Theoretical study of the deviation of any given state from the maximally entangled state using different
entanglement measures} \label{secII}} 

Consider a two-qubit pure state with Schmidt coefficients $c_0$ and $c_1$ as given below.
\begin{align}
\ket{\Psi} = c_{0}\ket{0}\ket{0}+c_1\ket{1}\ket{1}, 
\label{eq1}
\end{align}
where $c_{0}$ and $c_{1}$ satisfy the relation $0 \leq c_{0},c_{1} \leq 1 $ and $c_{0}^2 + c_{1}^2 = 1$. The state (\ref{eq1}) is maximally entangled for $c_{0},c_{1}=1/\sqrt{2}$, and  separable for $c_0, c_1 = 0,1$.
     
The three EMs discussed above for a two-qubit pure state are given by
\begin{subequations}
\begin{align}
\begin{split}
\text{N} &= c_{0}c_{1},
\end{split}\\
\begin{split}
\text{LN} &= \log_{2}(2c_{0}c_{1}+1),
\end{split}\\
\begin{split}
\text{EOF} &= - c_{0}^2\log_{2}c_{0}^2-c_{1}^2\log_{2}c_{1}^2.
\end{split}
\end{align}
\end{subequations}
    
In order to quantify the deviation of a given entangled state from the maximally entangled state, the following parameters are defined as measures of fractional deviations in terms of the quantities N, LN and EOF whose maximum values for the maximally entangled state are 0.5, 1, and 1, respectively.
\begin{subequations}
\begin{align}
\begin{split}
Q_{\text{N}} &= (0.5 - \text{N})/0.5~ ,  
\end{split}\\
\begin{split}
Q_{\text{L}} &= (1-\text{LN}),  
\end{split}\\
\begin{split}
Q_{\text{E}} &= (1 -\text{EOF}).
\end{split}\\
\end{align}
\end{subequations}

Note that all the above parameters range from 0 to 1, with 0 for the maximally entangled state and 1 for the separable state. For quantifying the extent to which these three parameters differ with each other, the following quantities are defined as absolute differences between the respective fractional deviations defined above.
\begin{subequations}
\begin{equation}
\Delta Q_{\text{NL}} = |Q_{\text{N}}- Q_{\text{L}}|, 
\end{equation}
\begin{equation}
\Delta Q_{\text{EL}} = |Q_{\text{E}}- Q_{\text{L}}|,
\end{equation}
\begin{equation}
\Delta Q_{\text{NE}} = |Q_{\text{N}}- Q_{\text{E}}|.
\end{equation}
\label{eq4}
\end{subequations}
    
Different values of the quantities $Q_{\text{N}}$, $Q_{\text{L}}$, $Q_{\text{E}}$, $\Delta Q_{\text{NL}}$, $\Delta Q_{\text{EL}}$, and $\Delta Q_{\text{NE}}$ corresponding to different values of Schmidt coefficients have been incorporated in Table~\ref{tab1} as percentage values. It is evident from Table~\ref{tab1} that for a given entangled state the percentage deviations are different for different EMs. For example, for a state with $c_{0}$ = 0.4 (where $c_{0}$ = 0.7071 corresponds to the maximally entangled state),  its percentage deviation from the maximally entangled state is 26.68\% when N is used to quantify entanglement; the percentage deviation is 20.66\% when LN is used to quantify entanglement, and is 36.57\% when one uses EOF. Thus, in this case, the differences in the percentage deviations are, respectively, given by  $\Delta Q_{\text{NL}}$ = 6.02\% , $\Delta Q_{\text{EL}}$ =  15.91\%, and $\Delta Q_{\text{NE}}$ = 9.89\%. Numerical study by optimization of the Schmidt coefficients to find the maximum deviations in $\Delta Q$ leads to the following results: \\
\begin{itemize}
\item Maximum value of $\Delta Q_{\text{NL}}$ is 8.61\% corresponding to the states with $c_{0}$ = 0.227  and 0.974.
\item Maximum value of $\Delta Q_{\text{EL}}$  is 23.57\% corresponding to the states with  $c_{0}$ = 0.217 and 0.976.
\item Maximum value of $\Delta Q_{\text{NE}}$  is 14.99\% corresponding to the states with $c_{0}$ = 0.210 and 0.978.
\end{itemize}

\begin{table}[htb]
\begin{center}
\resizebox{\textwidth}{!}{
\renewcommand{\arraystretch}{2}
\begin{tabular}{|l||c|c|c|c|c|c|c|c|c|r|}
\hline
$\mathbf{c_{0}}$ &  \textbf{N} & \textbf{LN} & \textbf{EOF} & $\mathbf{Q_{N}}(\%)$ & $\mathbf{Q_{L}}(\%)$ & $\mathbf{Q_{E}}(\%)$ &   $ \mathbf{\Delta Q_{NL}}(\%)$  & $\mathbf{\Delta Q_{EL}}(\%)$  & $\mathbf{\Delta Q_{NE}}(\%)$ \\
\hline \hline
0.1  &  0.099 & 0.262 &   0.081  & 80.10  & 73.82 & 91.92&  6.28 & 18.10  & 11.82 \\
\hline
0.2  &  0.196 & 0.477 &   0.242 & 60.81 & 52.29  & 75.77 & 8.52 & 23.48  & 14.96  \\
\hline
0.4  &  0.367 & 0.793 &   0.634 & 26.68  & 20.66  & 36.57   & 6.02  & 15.91  & 9.89  \\
\hline
0.7  &  0.499 & 0.999 &   0.999 & 0.02  & 0.01  &  0.03  &  0.01  & 0.01  & 0.01 \\
\hline
0.7071 &  0.5 & 1  & 1 & 0 & 0 & 0  &  0  & 0  & 0  \\
\hline
0.8  &  0.480 & 0.971 &   0.943 & 4.00  & 2.91  & 5.73  &  1.09  & 2.82  & 1.73 \\
\hline
0.9  &  0.392 & 0.836 &   0.701 & 21.54 & 16.44  & 29.85  & 5.10  & 13.41  & 8.31   \\
\hline
\end{tabular}}
\caption{\textit{Differences between the respective fractional deviation parameters for different EMs for two-qubit pure states given as percentage values.} \label{tab1}}
\end{center}
\end{table}
 
Note that the disagreement between the values of $\Delta Q$s increases for the states further away from the maximally entangled state, reaching a maximum value, and then decreases for the states getting closer to the separable state. Now, in order to analyze the way the above mentioned differences between the deviation parameters occur, we obtain the following results by studying the derivatives of different EMs with respect to the Schmidt coefficient  $c_{0}$ characterizing the two-qubit pure state.
     
The derivative of N with respect to $c_{0}$ is given by
\begin{align}
\dfrac{\text{dN}}{dc_{0}} = \frac{1-2c_{0}^2}{\sqrt{1-c_{0}^2}}.
\label{eq5}
\end{align}    
The derivative of LN with respect to $c_{0}$ is given by
\begin{align}
\dfrac{d\text{LN}}{dc_{0}} =  \frac{2(1-2c_{0}^2)}{\sqrt{1-c_{0}^2}~[2c_{0}\sqrt{1-c_{0}^2}+1]\ln(2)}.
\label{eq6}
\end{align}  
The derivative of EOF with respect to $c_{0}$ is given by
\begin{align}
\dfrac{d\text{EOF}}{dc_{0}} = \frac{2c_{0}\log_{2}[(1-c_{0}^2)/c_{0}^2]}{\ln(2)}.
\label{eq7}
\end{align}  

It can then be seen that the values of 2N, LN and EOF increase with $c_{0}$ starting from 0 and reach their respective maximum values corresponding to the maximally entangled state when $c_{0}$ is $1/\sqrt{2}$ and then start decreasing with further increasing values of $c_{0}$. Note that although the values of 2N, LN and EOF start from 0, the quantity LN kicks off rapidly due to the higher value of its derivative with respect to the state parameter $c_{0}$ as compared to that of 2N and EOF. Hence, for any $c_{0}$, the quantity LN is always greater than 2N and EOF, and has least deviation from the maximally entangled state. On the other hand, for any value of $c_{0}$, EOF is always less than 2N and LN,  and  has the highest deviation from the maximally entangled state. These features are illustrated in Figs.~\ref{fig2a}, \ref{fig2b} and Figs.~\ref{fig3a}, \ref{fig3b}. Monotonicity of N, EOF and LN with respect to each other can be further verified from Eqs.~(\ref{eq5}-\ref{eq7}) by computing the quantities $\frac{d\text{N}}{d\text{EOF}}$, etc.

\begin{figure}[H]
\centering
\subfigure[]{
\includegraphics[height=2.1in,keepaspectratio]{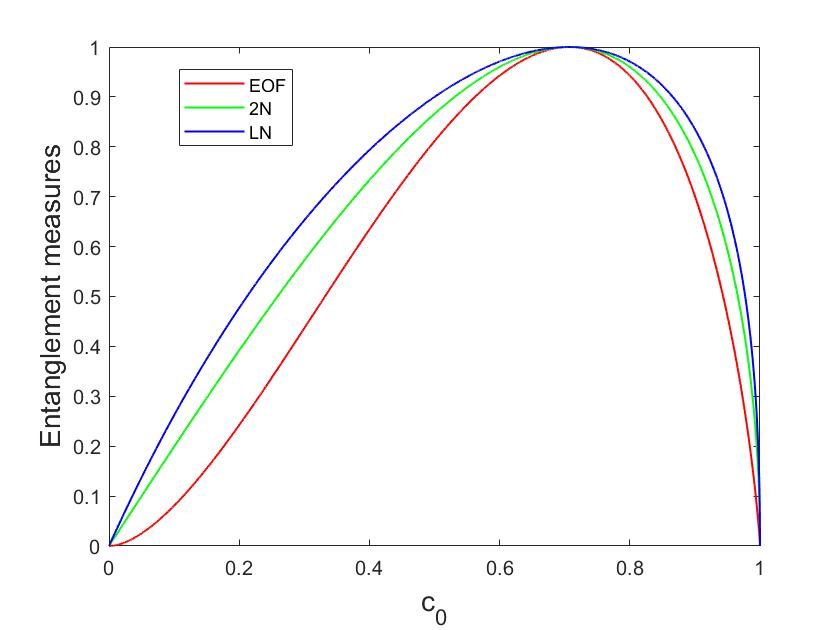}\label{fig2a}}
\subfigure[]{
\includegraphics[height=2.1in,keepaspectratio]{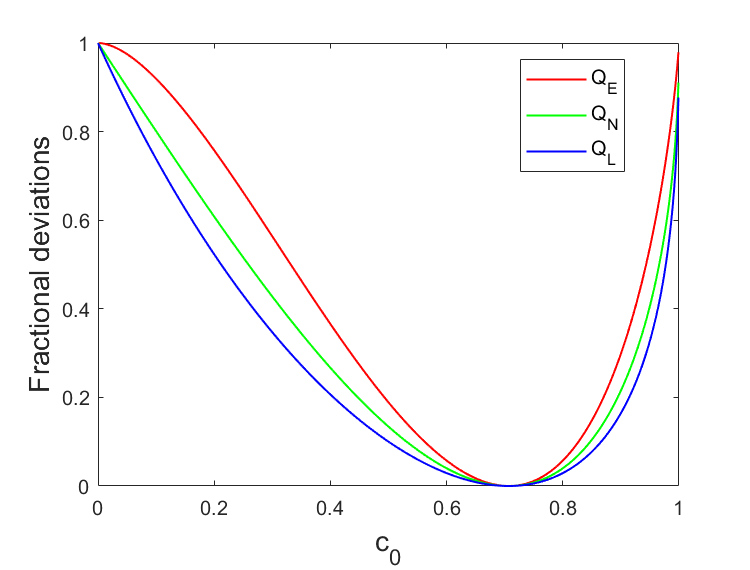}\label{fig2b}}
\caption{(a) \textit{This figure illustrates the variation of different entanglement measures with respect to the state parameter $c_{0}$. (b) This figure shows how the fractional deviations of a given state from the maximally entangled state calculated using different entanglement measures 2N, LN and EOF vary with respect to the state parameter $c_{0}$.}}
\end{figure}}

\begin{figure}[H]
\centering
\subfigure[]{
\includegraphics[height=2.2in,keepaspectratio]{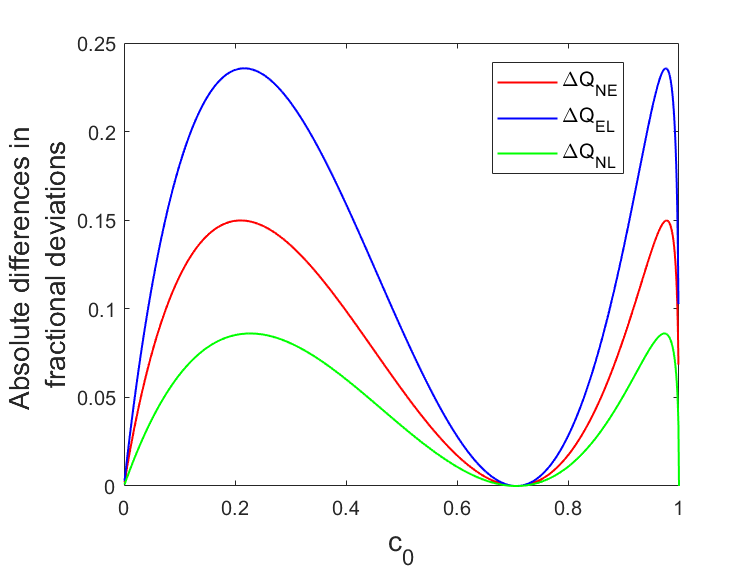}\label{fig3a}}
\subfigure[]{
\includegraphics[height=2.2in,keepaspectratio]{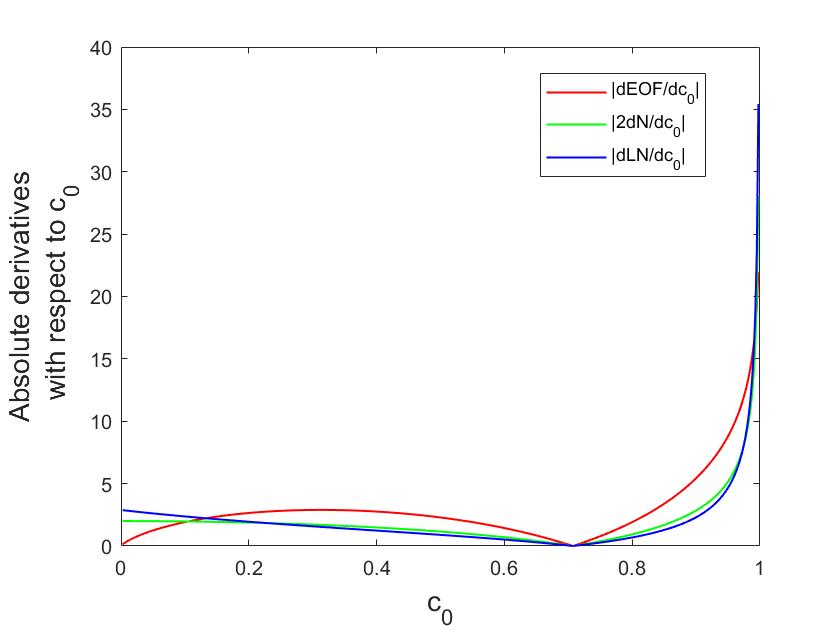}\label{fig3b}}
\caption{(a) \textit{This figure illustrates the differences in the fractional deviations of the respective states from the maximally entangled state calculated using different entanglement measures N, LN and EOF by varying the state parameter $c_{0}$. (b) This figure shows how the absolute values of  the derivatives of different entanglement measures 2N, LN and EOF with respect to state parameter $c_{0}$, vary with $c_{0}$.}}
\label{fig3}
\end{figure}
\textcolor{darkblue}{\section{Experimental study of the deviation of the prepared state from maximally entangled state using different entanglement measures} \label{secIII}}
 
Polarization entangled photon pairs are produced by the second order non-linear optical process of Spontaneous Parametric Down-Conversion (SPDC) in a two-crystal geometry [\ref{14}]. A 100 mW, Continuous Wave (CW) diode laser having central wavelength at 405 nm and a bandwidth of 1.2 nm (405 nm Cobolt-06-01-Series) was used as the pump laser. Two type-I BBO ($\beta-B_aB_2O_4$) crystals in sandwich configuration ($5\times 5\times 0.5 ~\text{mm}^3$ each from Castech Inc., China) having their optic axes orthogonal to each other and phase matched at $\theta=28.9^\circ$ and $\phi=0^\circ$ with half opening angle of the cone equal to $3^\circ$ was used for producing entangled photon pairs. Schematic of the experimental set up is shown in Fig.~\ref{fig4} below. 
 \begin{figure} [!htb]
\begin{center}
\includegraphics[clip, trim=4.5cm 16cm 3.95cm 2.5cm, width=0.7\linewidth]{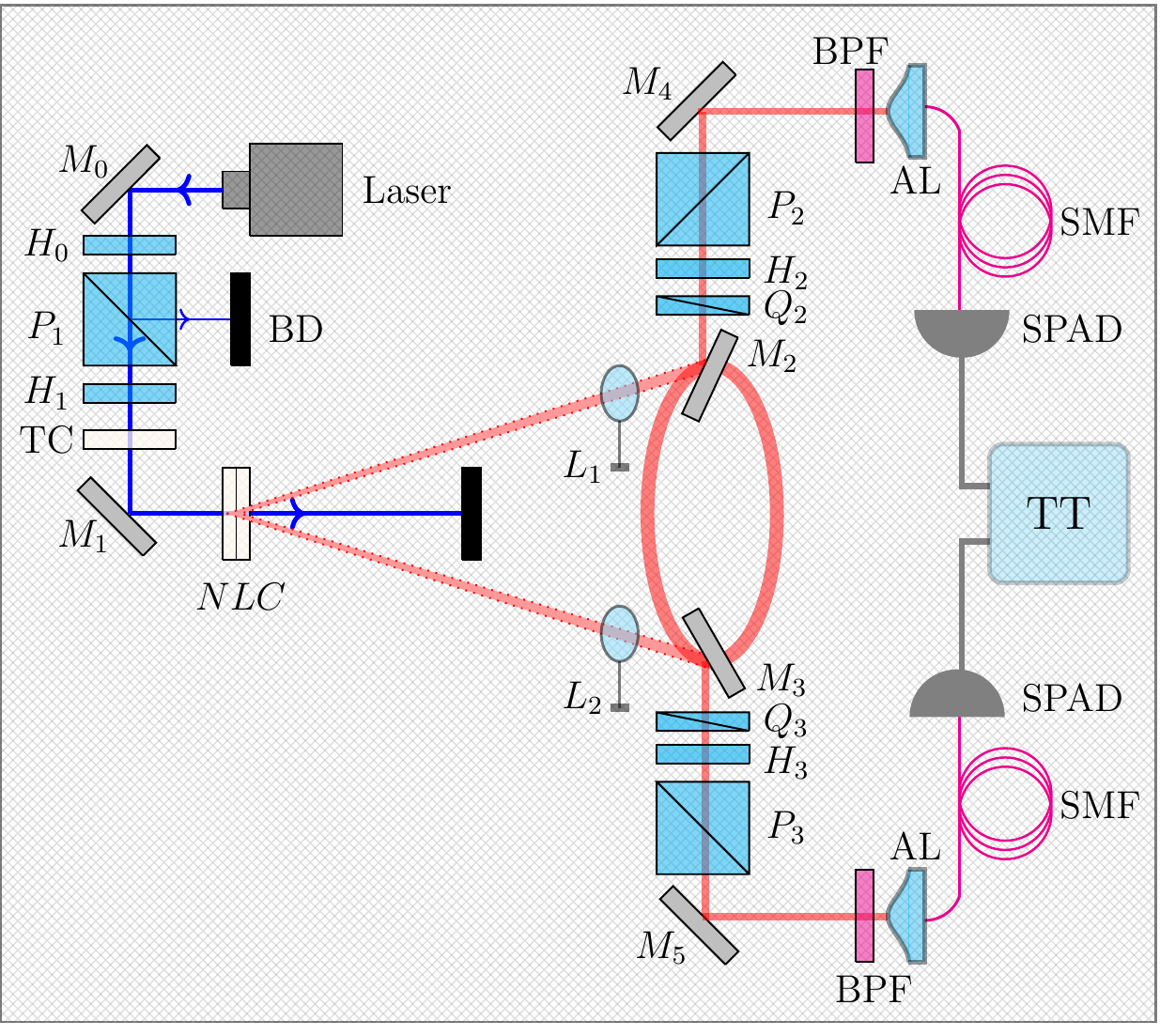}
\end{center}
\caption{\textit{Schematic of the experimental apparatus (not to scale) for preparation of SPDC based type-I polarization entangled photon source using two-crystal geometry and characterization using quantum state tomography. Different symbols have the following meaning: P, polarizing beam splitter; Q, quarter wave plate; H, half wave plate; NLC, non-linear crystal; TC, temporal compensator; L, plano-convex lens; M, mirror; BPF, bandpass filter; AL, aspheric lens; SMF, single mode fiber; SPAD, single-photon avalanche diode; and TT, time tagger unit or coincidence module.}}
\label{fig4}
\end{figure}

The pump beam is passed through a half-wave plate (HWP) and polarizing beam splitter (PBS) to get pure H-polarized laser beam. This laser beam is then passed through a HWP ($\text{H}_1$) with fast axis oriented at an angle $\theta$ with respect to vertical, which prepares the pump polarization state to be inputted to the BBO crystals for the preparation of different entangled states.

\begin{equation}
|H\rangle \xrightarrow{\text{HWP at} ~\theta} \sin(2\theta)|H\rangle+\cos(2\theta)|V\rangle =\alpha |H\rangle+\beta|V\rangle,
\end{equation}
where $\alpha=\sin(2\theta)$, and $\beta=\cos(2\theta)$.

Generation of entangled photons through SPDC process in a two-crystal geometry can be understood as follows: a H-polarized pump photon gets down-converted in the first crystal (whose optic axis is in horizontal plane) into a pair of V-polarized photons, and a V-polarized pump photon gets down-converted in the second crystal (whose optic axis is in vertical plane) into a pair of H-polarized photons. If these two processes occur in a coherent manner, i.e., they are indistinguishable, the two down-converted polarization amplitudes are coherently added and the resultant state becomes an entangled state as given below.
\begin{equation}
\begin{aligned}
\begin{split}
\alpha|H\rangle+\beta|V\rangle 
\xrightarrow {\text{SPDC}} \alpha |VV\rangle+\beta \exp(i\phi)|HH\rangle.
\end{split}
\end{aligned}
\end{equation}

The relative phase $\phi$ depends on the optical path difference/delay between the photons down-converted in the first and second crystals. It can be controlled by a tilting a quarter-wave plate in the pump beam (not shown in Fig.~\ref{fig4}).

The SPDC photons created in the first crystal get delayed compared to those created in the second crystal, thus giving rise to temporal distinguishability leading to drop in the quality of entanglement. This temporal delay is pre-compensated [\ref{15}] using another type-I BBO crystal (TC) of thickness 1.6 mm. The SPDC photons are then passed through a Quantum State Tomography (QST) setup [\ref{16}]  consisting of quarter-wave plate, half-wave plate and PBS on either side and collected through single mode fiber using aspheric lens and 810-10 nm band pass filter used for spectral filtering. These photons are then detected by single photon detectors and 36-coincidence measurements are performed for acquisition time of 60 s each. These measurements correspond to the projections in different bases that are required in QST for the state reconstruction. The maximum-likelihood estimation (MLE) [\ref{16}] is used to get the physical state (density matrix) from the QST data which is expected to have some experimental imperfections. 

Here, we have prepared three different two-qubit entangled states (three sets each) for pump HWP oriented at $23.1^\circ$ (State-I), $13.1^\circ$ (State-II) and $9.1^\circ$ (State-III). These states have average purity (where purity is denoted by P and defined as Tr[$\rho^2$], $\rho$ being the density matrix of the system) better than $95.7\%$. The representative 3D plots of the density matrices reconstructed through QST and MLE are shown in Figs.~\ref{fig5}-\ref{fig7} below.
\begin{figure}[H]
\centering
\includegraphics[width=0.6\columnwidth, keepaspectratio]{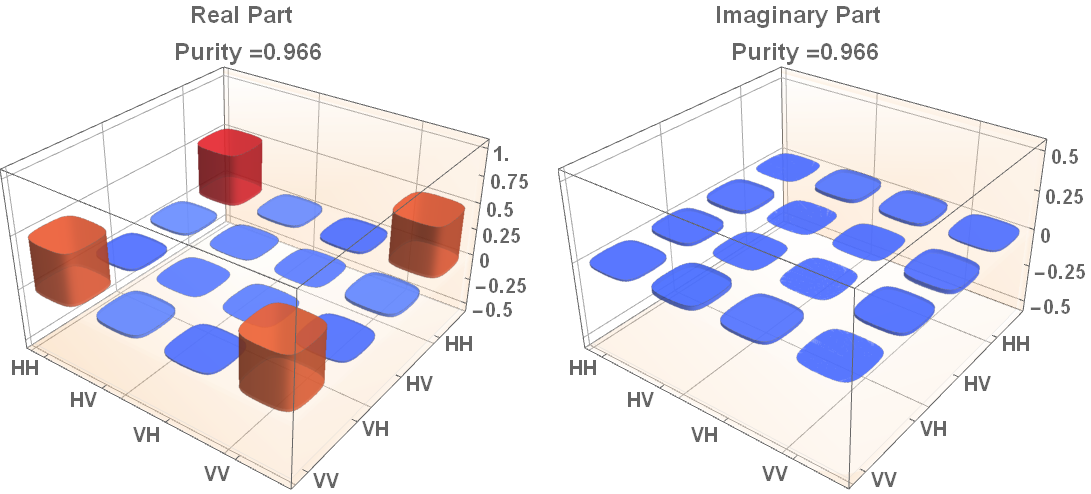}
\caption{\textit{Representative 3D plot of the experimentally reconstructed density matrix for state-I with P=0.966 and 2N=0.964.}}
\label{fig5}
\end{figure}

\begin{figure}[H]
\centering
\includegraphics[width=0.6\columnwidth, keepaspectratio]{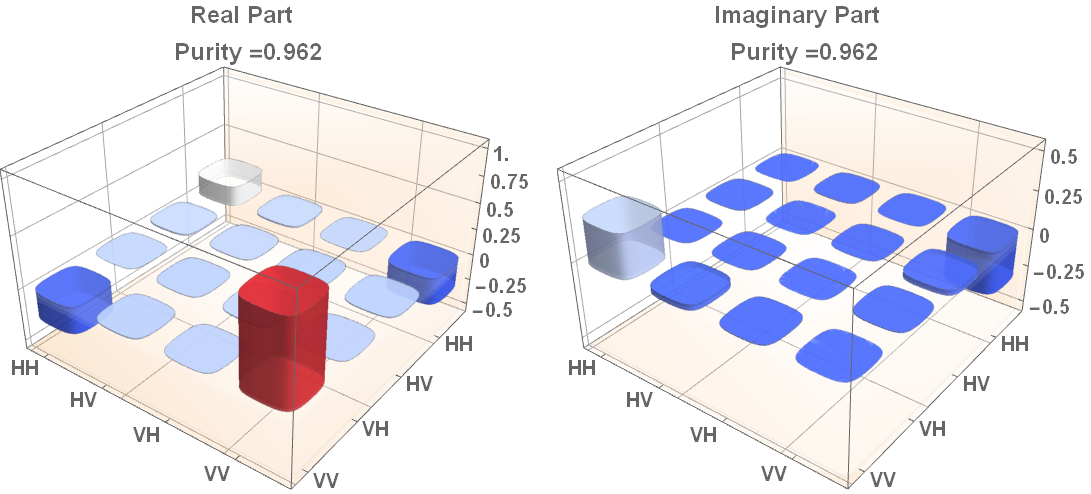}
\caption{\textit{Representative 3D plot of the experimentally reconstructed density matrix for state-II with P=0.962 and 2N=0.759.}}
\label{fig6}
\end{figure}

\begin{figure}[!htb]
\centering
\includegraphics[width=0.6\columnwidth, keepaspectratio]{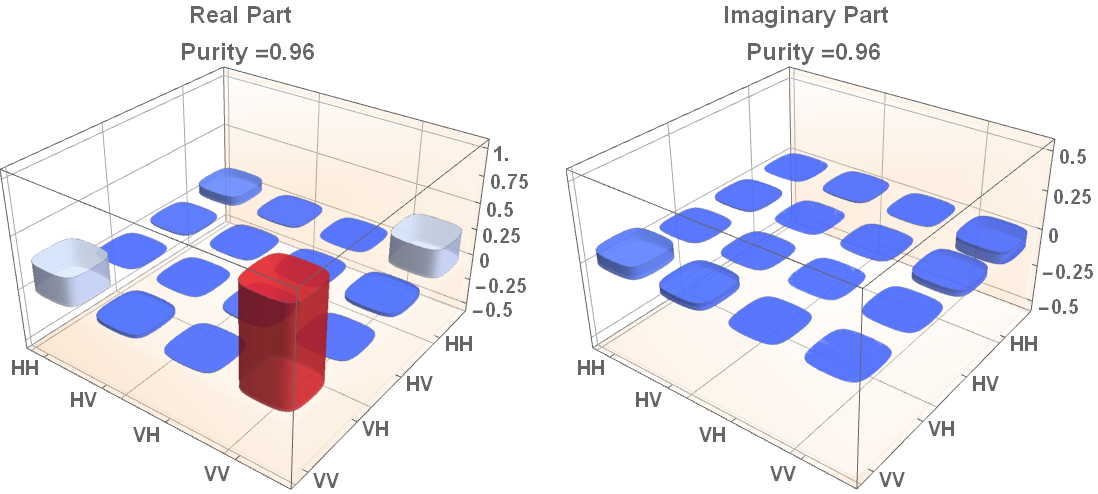}
\caption{\textit{Representative 3D plot of the experimentally reconstructed density matrix for state-III with P=0.960 and 2N=0.544.}}
\label{fig7}
\end{figure}

Properties of different experimentally prepared entangled states such as purity, quantification of entanglement by different EMs, and their respective deviations from the maximally entangled state are summarized in the Table~\ref{tab2} below. The statistical error due to reconstruction occurs in the third decimal place (indicated in parentheses in the Table~\ref{tab2}) for P, 2N, LN and EOF. Thus statistical errors in other derived quantities would also be of the same order.
\begin{table}[!htb] 
\begin{center}
\resizebox{\textwidth}{!}{
\renewcommand{\arraystretch}{2}
\begin{tabular}{|l||c|c|c|c|c|c|c|c|c|c|c|c|c|r|}
\hline
\textbf{States} & $\mathbf{2N_{ideal}}$ &   $\mathbf{P_{expt}}$ & $\mathbf{2N_{expt}}$ & $\mathbf{LN_{expt}}$ & $\mathbf{EOF_{expt}}$  &  $\mathbf{Q_L (\%)}$  &  $\mathbf{Q_N (\%)}$  &  $\mathbf{Q_E (\%)}$  & $\mathbf{\Delta Q_{NL}(\%)}$  & $\mathbf{\Delta Q_{NE}(\%)}$ & $\mathbf{\Delta Q_{EL}(\%)}$  \\
\hline
\hline
I  & 0.999 &   0.961(5) & 0 0.960(5) & 0.971(4) & 0.944(6) & 4.01 & 2.93 & 5.57 & 1.09  & 1.55 & 2.64\\
\hline
II  & 0.792 &  0.957(4) &  0.749(9) & 0.808(7) & 0.663(12) & 25.07 & 19.32 & 33.68 & 5.75 & 8.61 & 14.36 \\
\hline
III  & 0.593 & 0.958(2)   &  0.547(3) & 0.629(3) & 0.413(5) & 45.32 & 37.07 & 58.70 & 8.25  & 13.39 & 21.64\\
\hline
\end{tabular}}
\caption{\textit{Comparison tables for the properties of experimentally prepared two-qubit entangled states and its deviation from the intended maximally entangled state. Different Q-values are reported as percentage quantity.}\label{tab2}}
\end{center} 
\end{table}

It is evident from the experimentally prepared states considered here that $\Delta Q_{\text{EL}}$  $>$ $\Delta Q_{\text{NE}}$ $> $ $\Delta Q_{\text{NL}}$. Further, for the State-III, the values of $\Delta Q_{\text{NL}}$,  $\Delta Q_{\text{EL}}$, and  $\Delta Q_{\text{NE}}$ are very close to the maximum deviation obtained by numerical optimization, albeit with small impurity in the experimental states. These observations are in close agreement with the expectations from the analytical derivations that have been shown in the theory Section~\ref{secII}. 

\textcolor{darkblue}{\section{Extending the analysis to higher-dimensional systems}\label{secIV}}

In this section, we will discuss the way the present analysis can be extended to higher dimensional systems by considering bipartite pure qutrit states and multipartite states such as tripartite pure qubit states.

\textcolor{darkblue}{\subsection{Bipartite pure qutrit states}}

Consider a two-qutrit pure state with Schmidt coefficients $c_{0}$, $c_{1}$, and $c_{2}$ as given below.
\begin{align}
\ket{\Phi} = c_{0}\ket{0}\ket{0} + c_{1}\ket{1}\ket{1} + c_{2}\ket{2}\ket{2},
\label{eq10}
\end{align}
where $0 \leq c_{0},~c_{1},~c_{2} \leq 1$ and $c_{0}^2$ + $c_{1}^2$  + $c_{2}^2$ = 1. The state (\ref{eq10}) is maximally entangled for $c_{0}$ = $c_{1}$ = $c_{2}$ = 1/$\sqrt{3}$ and separable when one of the Schmidt coefficients is one and others are zero. 

We will first recap some of the well known entanglement measures for higher dimensional systems. Negativity ($\mathcal{N}$) [\ref{17}], Entanglement of formation ($\mathcal{E}$) [\ref{18}], normalized generalized Concurrence ($\mathcal{C}$) [\ref{19}], and Linear entropy ($\mathcal{I}$) [\ref{20}] for two-qutrit  pure states are defined as follows:
 \begin{subequations}
\begin{align}
\begin{split}
\mathcal{N} &= c_{0}c_{1} + c_{1}c_{2} + c_{2}c_{0},
\end{split} \\
\begin{split}
\mathcal{E} &= -c_{0}^2(log_{2}c_{0}^2) - c_{1}^2(log_{2}c_{1}^2) - c_{2}^2(log_{2}c_{2}^2),
\end{split} \\
\begin{split}
\mathcal{C} &= \sqrt{3(c_{0}^2c_{1}^2+c_{1}^2c_{2}^2+c_{2}^2c_{0}^2)},
\label{eq3.11c}
\end{split} \\
\begin{split}
\mathcal{I} &= 3(c_{0}^2c_{1}^2+c_{1}^2c{2}^2+c_{2}^2c_{0}^2).
\label{eq3.11d}
\end{split} 
\end{align}
\end{subequations}

From Eq.~(\ref{eq3.11c}) and (\ref{eq3.11d}), it is clear that the Linear entropy is just the square of generalized Concurrence. Hence, we shall consider generalized Concurrence and compare it with other EMs. In order to quantify the percentage deviation of a given non-maximally entangled two-qutrit pure state from the maximally entangled state using different EMs, we define the Q-parameters as given below.
\begin{subequations}
\begin{align}
\begin{split}
Q_{\mathcal{E}} = \frac{\log_{2}(3) - \mathcal{E}}{\log_{2}(3)},
\end{split}\\
\begin{split}
 Q_{\mathcal{N}} = (1 - \mathcal{N}),
\end{split}\\
\begin{split}
Q_{\mathcal{C}} = (1 - \mathcal{C}).
\end{split}
\end{align}
\end{subequations}

In this context, in a recent earlier work from our group [\ref{21}], extensive theoretical as well as experimental analysis on the certification and quantification of entanglement in spatial-bin bipartite photonic qutrits was done. Here, significant differences between the computed deviations of any pure non-maximally entangled bipartite qutrit state from the maximally entangled state in terms of the EMs such as $\mathcal{E}$ and $\mathcal{N}$ were reported for a different pair of Schmidt coefficients from the ones reported here. Further, it was shown that $\mathcal{E}$ and $\mathcal{N}$ are not monotonic with respect to each other. In this work, we consider, in addition to the comparison between $\mathcal{E}$ and $\mathcal{N}$, the entanglement measure $\mathcal{C}$ as well.

In order to quantify to what extent these three parameters differ with each other, the following quantities are appropriate measures.
\begin{subequations}
\begin{align}
\begin{split}
\Delta Q_{\mathcal{NE}} &= |Q_{\mathcal{N}}- Q_{\mathcal{E}}|~,
\end{split}\\
\begin{split}
\Delta Q_{\mathcal{EC}} &= |Q_{\mathcal{E}}- Q_{\mathcal{C}}|~,
\end{split}\\
\begin{split}
\Delta Q_{\mathcal{NC}} &= |Q_{\mathcal{N}}- Q_{\mathcal{C}}|~.
\end{split}
\end{align}
\end{subequations}

Now, we will present some specific cases exemplifying the non-monotonic nature of these EMs.
\begin{itemize}
\item Consider a state $\ket{\phi_{1}}$ with Schmidt coefficients $c_{0}$ = 0.9755 and $c_{1}$ = 0.0361, and another state $\ket{\phi_{2}}$ with Schmidt coefficients $c_{0}$=0.1403 and $c_{1}$ = 0.1346. 
    $$ \mathcal{E}(\phi_{1}) = 0.2878\  ~\text{and}~ \  \mathcal{N}(\phi_{1}) = 0.2546~,$$
    $$ \mathcal{E}(\phi_{2}) = 0.2698\  ~\text{and}~  \mathcal{N}(\phi_{2}) = 0.2885~.$$
Here, $\mathcal{E}(\phi_{1}) > \mathcal{E}(\phi_{2})$ but $\mathcal{N}(\phi_{1}) < \mathcal{N}(\phi_{2})$, showing that $\mathcal{E}$ and $\mathcal{N}$ are not monotonic with respect to each other.

\item Consider a state $\ket{\phi_{3}}$ with Schmidt coefficients $c_{0}$ = 0.4134 and $c_{1}$ = 0.8275, and another state $\ket{\phi_{4}}$ with Schmidt coefficients $c_{0}$ = 0.7452 and $c_{1}$ = 0.1143. 
    $$ \mathcal{N}(\phi_{3}) = 0.8136\  ~\text{and}~  \mathcal{C}(\phi_{3}) = 0.8495~, $$
    $$ \mathcal{N}(\phi_{4}) = 0.6498 \  ~\text{and}~      \mathcal{C}(\phi_{4}) = 0.8705~. $$
Here, $\mathcal{C}(\phi_{3}) < \mathcal{C}(\phi_{4})$ but $\mathcal{N}(\phi_{3}) > \mathcal{N}(\phi_{4})$, showing that $\mathcal{C}$ and $\mathcal{N}$ are not monotonic with respect to each other.

\item Consider a state $\ket{\phi_{5}}$ with Schmidt coefficients $c_{0}$ = 0.4134 and $c_{1}$ = 0.8275 and another state $\ket{\phi_{6}}$ with Schmidt coefficients $c_{0}$ = 0.2334 and $c_{1}$ = 0.8052. 
    $$ \mathcal{E}(\phi_{5}) = 1.2128\  ~\text{and}~  \mathcal{C}(\phi_{5}) = 0.8495~,$$
    $$ \mathcal{E}(\phi_{6}) = 1.1542 \  ~\text{and}~  \mathcal{C}(\phi_{6}) = 0.8559~.$$
Here, $\mathcal{C}(\phi_{5}) <\mathcal{C}(\phi_{6})$ but $\mathcal{E}(\phi_{5}) > \mathcal{E}(\phi_{6})$, showing that $\mathcal{C}$ and $\mathcal{E}$ are not monotonic with respect to each other.
\end{itemize}

Different values of the quantities $Q_{\mathcal{E}}$, $Q_{\mathcal{N}}$, $Q_{\mathcal{C}}$, $\Delta Q_{\mathcal{NE}}$, $\Delta Q_{\mathcal{EC}}$, and $\Delta Q_{\mathcal{NC}}$ corresponding to different values of Schmidt coefficients have been incorporated in Table~\ref{tab3} as percentage values.

\begin{table}[!h]
\begin{center}
\resizebox{\textwidth}{!}{
\renewcommand{\arraystretch}{2}
\begin{tabular}{|l|c||c|c|c|c|c|c|c|c|r|}
\hline
\textbf{$c_{0}$} & \textbf{$c_{1}$} & $\mathcal{E}$ & $\mathcal{N}$ & $\mathcal{C}$ & $Q_\mathcal{E}$ & $Q_\mathcal{N}$  & $Q_\mathcal{C}$ & $\Delta Q_\mathcal{NE}$ & $\Delta Q_\mathcal{EC}$ & $\Delta Q_\mathcal{NC}$\\
\hline \hline
0.1 & 0.1 & 0.1614 & 0.2080 & 0.2431 & 89.81 & 79.20 & 75.69 & 10.61 & 14.12 & 3.51 \\
\hline
0.3 & 0.8 & 1.2347 & 0.8116 & 0.8741 &  22.10 & 18.84 & 12.59 & 3.25 & 9.51 & 6.26 \\
\hline
0.5774 & 0.5774 & 1.5850 & 1 & 1 & 0 & 0 & 0 & 0 & 0 & 0 \\
\hline
0.6 & 0.6 & 1.5755  & 0.9950 & 0.9968  & 0.60& 0.50 & 0.32 & 0.10 & 0.28 & 0.18 \\
\hline
0.9 & 0.3 & 0.8911 & 0.6495 & 0.6990 &  43.78 & 35.05 & 30.09 & 8.73 & 13.69 & 4.96\\
\hline
\end{tabular}}
\caption{\textit{Differences in the percentage deviations of entanglement measures of a given state from the value corresponding to the maximally entangled two-qutrit pure state.} \label{tab3}}
\end{center}
\end{table}

\textcolor{darkblue}{\textbf{Observations:}}
\begin{itemize}
\item For non-maximally entangled two-qutrit pure states, a given EM is not always greater (or less) than any other EM for different values of the state parameters, unlike for the case of two-qubit pure states.

\item From numerical optimization, we find that the $\Delta Q_{\mathcal{NE}}$ takes a maximum value of 13.09\% when $c_{0} = 0.7071$ and  $c_{1} = 0.7071$, $\Delta Q_{\mathcal{EC}}$ takes a maximum value of 23.81\% when $c_{0} = 0.5$  and  $c_{1} = 0.8660$, and $\Delta Q_{\mathcal{NC}}$ takes a maximum value of 36.60\% when $c_{0} = 0.7071$ and  $c_{1} = 0.7071$.
\end{itemize}

We would like to emphasize an important and interesting point that although in the bipartite qubit case different EMs show different deviations of a given state from maximally entangled state, the EMs are monotonic with respect to each other. But in the bipartite qutrit case, different EMs not only provide different estimations of the deviation of any non-maximally entangled state from the maximally entangled state, the EMs can also be non-monotonic with respect to each other.

\textcolor{darkblue}{\subsection{Tripartite pure qubit states}}
Here, we indicate the directions for extending the analysis presented in this paper to multipartite systems focusing primarily on tripartite pure qubit states. Let us consider the usually discussed tripartite pure qubit states such as GHZ-type state [\ref{22}], W-state [\ref{23}] and Cluster state [\ref{24}]. In tripartite system, Cluster state is same as the GHZ state [\ref{24}]. Let us consider the following tripartite pure qubit state
\begin{align}
\ket{\xi_1} = c_{0}\ket{000} + c_{1}\ket{111},
\label{eq21}
\end{align}  
where $0 \leq c_{0}, c_{1} \leq 1$ and $c_{0}^2 + c_{1}^2 = 1$. The state (\ref{eq21}) is precisely the GHZ-state for $c_{0}= c_{1} = 1/\sqrt{2}$ when it is maximally entangled.

We consider two of the most commonly discussed EMs in tripartite system. One is the Tangle introduced by Coffman \textit{et al.}  [\ref{25}] which can be regarded as a hyperdeterminant of second order [\ref{26}]. Tangle for the state (\ref{eq21}) is given by
\begin{align}
\tau = 4c_{0}^2c_{1}^2 .
\end{align}

Another relevant EM is global measure of entanglement($G$) introduced by Meyer \textit{et al.} [\ref{27}] in Brennon form [\ref{28}-\ref{30}]. The value of this measure for the state~(\ref{eq21}) is given by
\begin{align}
G = 4c_{0}^2c_{1}^2.
\end{align} 

It is found that the expressions for EMs $G$ and $\tau$ are same for the class of states given by Eq.~(\ref{eq21}). Hence, estimation of the deviation of any given entangled state of form~(\ref{eq21}) from the maximally entangled state using these two EMs will be the same, while both these measures give the same value. Both these measures give the same value one [\ref{25}, \ref{27}] for the GHZ state.

Next, we consider another commonly discussed tripartite pure qubit state of the form 
\begin{align}
\ket{\xi_2} = c_{0}\ket{001} + c_{1}\ket{010} + c_{2}\ket{100},  
\label{eq24}
\end{align}
where $0 \leq c_{0}, c_{1}, c_{2} \leq 1$ and $c_{0}^2$ + $c_{1}^2$ + $c_{2}^2$ = 1. The above state is a W-state for $c_{0}$ = $c_{1}$ = $c_{2}$ = 1/$\sqrt{3}$ when it is maximally entangled. 

On calculation, it is found that for the state~(\ref{eq24}) Tangle vanishes for all the values of the state parameters, whereas the other measure $G$ is non-zero and given by 
\begin{align}
 G = 8/3(c_{0}^2c_{1}^2+c_{1}^2c_{2}^2+c_{2}^2c_{0}^2).
\end{align}
 
It can be seen from above that for W-state, $G = 8/9$ whereas $\tau$ remains zero [\ref{25}-\ref{27}]. Hence, the deviation of any non-maximally entangled state of the form~(\ref{eq24}) from the maximally entangled W-state can only be estimated in terms of $G$, not using tangle. A curious observation is that although the two entanglement measures $\tau$ and $G$ are defined differently, but for the particular class of state in~(\ref{eq21}), they capture same amount of entanglement as well as give same amount of deviation from the maximally entangled state. This example is in stark contrast to the two-qubit case where different EMs give different deviations. On the other hand, for state~(\ref{eq24}), $\tau$ vanishes whereas $G$ remains non-zero. This highlights the importance of this enterprise of comparing and contrasting different EMs. It would be worth further developing this line of study using other EMs as well for the tripartite qubit case.

\textcolor{darkblue}{\section{Outlook}\label{secV}}

The appreciable quantitative disagreement between estimates of the percentage deviation of a given state from the maximally entangled state using different EMs, thoroughly shown in this paper using both theoretical and experiment-based (for bipartite qubit states) analyses, underscores the need for an appropriate quantifier for addressing such an empirically relevant issue which is of significance for evaluating the efficacy of a given entangled state for its various applications. In this context, it is relevant to note that the question of quantifying `distance' or deviation of a given state from the separable state has been raised earlier and for this purpose the concept of `distance measures'  has been suggested in two different ways; one of which is in terms of relative entropy of entanglement [\ref{04}] and the other using the notion of `robustness' [\ref{31}] in terms of the noise that is required to be added to a given state to make it a separable state. Taking clue from such studies, one line of study may be to formulate a suitable `distance measure' for capturing the departure of a given non-maximally entangled state from the maximally entangled state and compare the results obtained with the relevant estimates using different EMs.  On the other hand, one can take cue from the concept of `teleportation distance'  [\ref{10}] that has been used to quantify the degree of performance of the resource channel used for teleportation, where distance of the teleported state from the target state has been quantified in terms of the trace norm (T = $\mathrm{Tr}[\sqrt{A^{\dagger}A}]$). Adapting this measure, by using the Frobenius norm  [\ref{32}], one may invoke the following measure to signify how close a given state ($\psi$) is to the intended maximally entangled state ($\psi_{\text{max}}$) (the correct choice of the maximally entangled state $\psi_{\text{max}}$ is the one which has minimum distance from the given state)
\begin{align}
D = || \ket{\psi}\bra{\psi}-\ket{\psi_{\text{max}}}\bra{\psi_{\text{max}}}||~,
\end{align}

where the Frobenius norm for a density matrix $A$, $||A||$ used above is given by $\sqrt{\mathrm{Tr}[ A^{\dagger}A]}$. Here, the Frobenius norm is preferred over trace norm in order to ensure that the parameter $D$ ranges from 0 to 1 for two-qubit pure states, with $D=0$ for the maximally entangled state and $D=1$ for the separable state.

Another possible approach is in terms of the notion of fidelity  [\ref{33}] as a measure of `closeness' between two pure states. The metric defined using this notion of fidelity like Bures distance [\ref{34}, \ref{35}], or the Fubini-Study metric used by Anandan and Aharonov [\ref{36}] does not range from 0 to 1 for two-qubit pure states and hence is not useful for comparison with the results obtained using the deviation parameters (ranging from 0 to 1) defined in Section~\ref{secII}. On the other hand, by appropriately scaling the metric defined in terms of fidelity by Gilchrist \textit{et al.}, [\ref{37}], the following measure `$C$' which ranges from 0 to 1 for two-qubit pure states can be used for the purpose of comparison with the results obtained from the deviation parameters defined in terms of different EMs.
\begin{align}
C = \sqrt{2-2(\bra{\psi}\ket{\psi_{\text{max}}})^2}~,
\end{align}

where $\psi_{\text{max}}$ is the intended maximally entangled two-qubit pure state. Note that $C=0$ for the maximally entangled state and $C=1$ for the separable state. Now, we note that for any given $\ket{\psi}$, the two measures $D$ and $C$ defined above, can be shown analytically to be equivalent. The numerical study in this case by varying values of the state parameter $c_{0}$ in Eq.(\ref{eq1}), interestingly, shows that the above mentioned distance measure $D(C)$  provides  an upper bound to the fractional deviations of various states from the maximally entangled state, calculated using N and LN; i.e.,
\begin{align}
D(C) \geq Q_{\text{N}}, Q_{\text{L}}.
\end{align}

\begin{figure}[H]
\centering
\includegraphics[width = 0.7\linewidth]{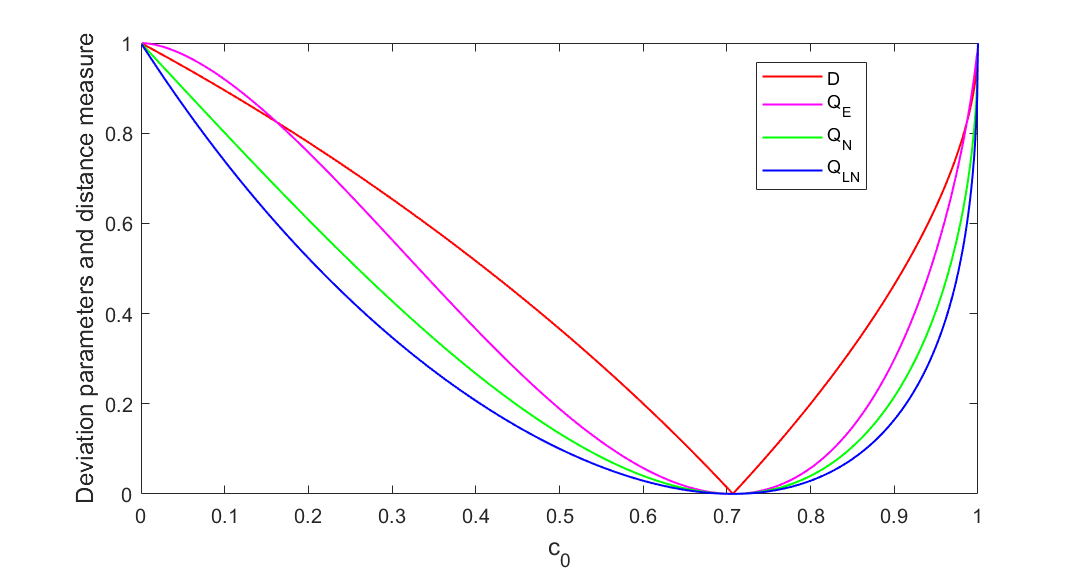}
\caption {\textit{ This figure illustrates that the curve for the distance measure $D$ provides an upper bound to the fractional deviation curves for N and LN, but not for EOF, since there are certain states for which the fractional deviation curve for EOF is not upper bounded by the curve related to $D$.}}
 \label{fig8}
\end{figure}

However,  there are certain states for which the respective fractional deviation from the maximally entangled state in terms of EOF, i.e., the quantity $Q_{\text{E}}$ is greater than $D(C)$, thus, restricting the use of $D(C)$ as an upper bound of such fractional deviations using EOF. These features pertaining to the distance measure $D(C)$ from the maximally entangled state are illustrated in Fig.~\ref{fig8}, whose implications may further be studied.

 A line of study different from the above mentioned approaches in terms of `distance measures' could be from the operational perspective of the use of quantum entanglement as resource; in other words, one may try to assess how close a non-maximally entangled state is to the maximally entangled state in terms of how useful it is as a resource. For example, in the context of demonstrating non-locality, in order to quantify the fractional deviation of a given non-maximally entangled state from the maximally entangled state, one may compute the maximum possible quantum mechanical violation of the Bell-CHSH inequality ($B_{V}$), i.e., the difference between the maximum quantum mechanical value of the Bell-CHSH expression for the given state and the lower bound of the Bell-CHSH expression. One can then estimate the fractional deviation of the parameter $B_{V}$ from its maximum value $(B_{V})_{\text{max}}$  = $2\sqrt{2}-2$ which is achieved for the maximally entangled state. Note that for any the  two-qubit pure state characterized by the Schmidt coefficients $c_{0}$ and $c_{1}$, $B_{V}$ for the two outcomes-two settings scenario is given by [\ref{38}-\ref{40}]
\begin{align}
B_{V} = 2\sqrt{1+(2c_{0}c_{1})^2}-2. 
\end{align} 

Now, similar to the other fractional deviation parameters defined in Section~\ref{secII}, here we define the following parameter as a measure of the fractional deviation of $B_{V}$ from its value $(B_{V})_\text{max}$ for the maximally entangled state.
\begin{align}
Q_{B_{V}} = \frac{(B_{V})_\text{max}-B_{V}}{(B_{V})_\text{max}}.
\end{align}

\begin{figure}[H]
\centering
\includegraphics[width =0.7\linewidth]{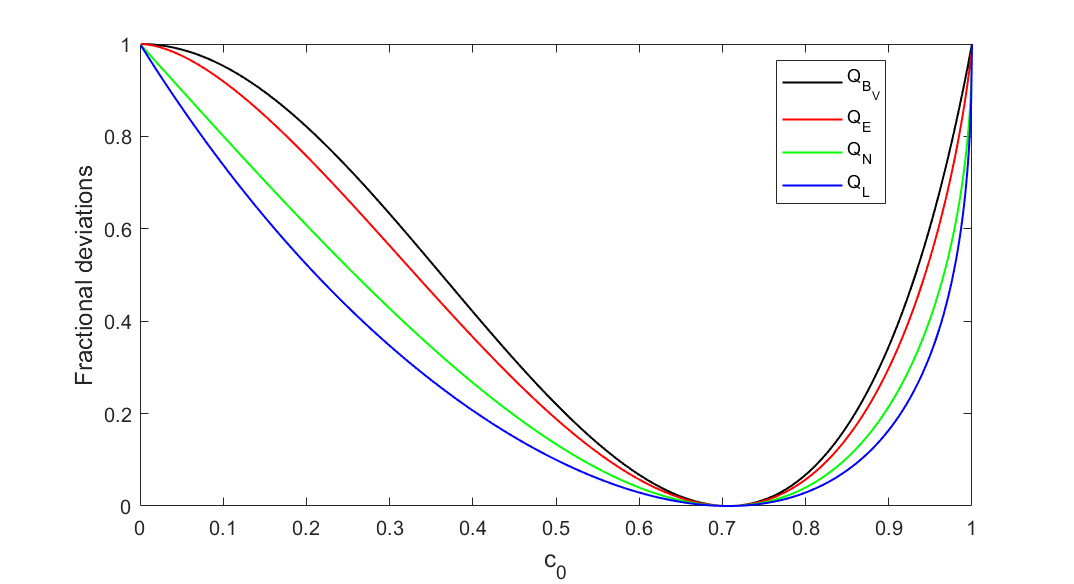}
\caption{{\textit{This figure shows that the fractional deviation curve for the quantity $B_{V}$ provides an upper bound to the other fractional deviation curves corresponding to the quantities N, LN and EOF, signifying that in terms of the amount of  Bell-CHSH violation, any given state is further away from the maximally entangled state than that estimated using different entanglement measures.}}} 
\label{fig9}
\end{figure}

Note that the values of $Q_{{B}_{V}}$ range from 0 to 1, with 0 for the maximally entangled state and 1 for the separable state. A numerical study of $Q_{B_{V}}$ by varying  values of the state parameter $c_{0}$ shows an interesting result that for any non-maximally entangled state, this fractional deviation parameter is greater than all other such fractional deviation parameters evaluated using different EMs like N, LN and EOF; i.e.,
\begin{align}
Q_{B_{V}} \geq Q_\text{N}, Q_\text{L},Q_\text{E},
\end{align}

where the equality holds good for the maximally entangled state and the separable state. This means that the measure of deviation of a given non-maximally entangled state from the maximally entangled state as quantified by the parameter $Q_{{B}_V}$ is always greater than that obtained from different EMs. For example, for a state with $c_{0}$ = 0.4, the fractional deviation of $B_{V}$ from its value corresponding to the maximally entangled state in percentage is 42.06\%; while the fractional deviations of N, LN and EOF from their values corresponding to the maximally entangled state in percentages are 26.68\%, 20.66\% and 36.57\%, respectively.  Illustration of this feature provided in Fig.~\ref{fig9}, thus, suggests nuances in the quantitative relationship between EMs and the amount of non-locality shown by the Bell-CHSH violation present even in the simplest two outcomes - two settings scenario involving two-qubit pure states; on the other hand, aspects of quantitative non-equivalence between entanglement and non-locality have so far been discussed essentially for high dimensional systems or scenarios involving larger number of settings [\ref{41}-\ref{44}].

\textcolor{darkblue}{\section{Concluding Remarks}\label{secVI}}

In sum, the results of studies of the present paper bring out the need for exploring different ideas for quantifying how close (far) a given entangled state is to (from) the maximally entangled state and comparing the results obtained by using such quantifiers with that based on different entanglement measures. This leads to the following question: Is there any fundamental criterion for assessing which quantifier is the appropriate one to be used for addressing questions such as the one posed in this paper, or whether such a criterion would have to be operationally defined essentially dependent on the specific context in which the entangled state is used as a resource? A comprehensive study is required for shedding further light on this issue as well as for gaining a deeper understanding of the comparison between different entanglement measures taking into account the studies probing their respective physical significance [\ref{02},\ref{03},\ref{45}]. 

\textcolor{darkblue}{\section{Acknowledgments}}
\textit{DH} thanks NASI for the support provided by Senior Scientist Platinum Jubilee Fellowship. \textit{IA} acknowledges the support by IISER, Mohali and DST INSPIRE-SHE programme in carrying out part of this work as a summer intern at Bose Institute, Kolkata. \textit{US} acknowledges partial support from the QuEST-DST network programme of the Govt. of India.

\textcolor{darkblue}{\section{References}}
\begin{enumerate}
\setlength{\itemsep}{0pt}
\item \textit{C. H. Bennett, H. J. Bernstein, S. Popescu, and B. Schumacher, ``Concentrating partial entanglement by local operations," \href{https://doi.org/10.1103/PhysRevA.53.2046}{Phys. Rev. A \textbf{53}, 2046 (1996)}}. \label{01}
\item \textit{C. H. Bennett, D. P. DiVincenzo, J. A. Smolin, and W. K. Wootters, ``Mixed-state entanglement and quantum error correction," \href{https://doi.org/10.1103/PhysRevA.54.3824}{Phys. Rev. A \textbf{54}, 3824 (1996)}}.  \label{02}
\item \textit{S. Popescu and D. Rohrlich, ``Thermodynamics and the measure of entanglement," \href{https://doi.org/10.1103/PhysRevA.56.R3319}{Phys. Rev. A \textbf{56}, R3319 (1997)}}.  \label{03}
\item \textit{V. Vedral, M. B. Plenio, M. A. Rippin, and P. L. Knight, ``Quantifying entanglement," \href{https://doi.org/10.1103/PhysRevLett.78.2275}{Phys. Rev. Lett. \textbf{78}, 2275 (1997)}}.\label{04}
\item \textit{V. Vedral and M. B. Plenio, ``Entanglement measures and purification procedures," \href{https://doi.org/10.1103/PhysRevA.57.1619}{Phys. Rev. A \textbf{57}, 1619 (1998)}}.\label{05}
\item \textit{G. Vidal, ``Entanglement monotones," \href{https://doi.org/10.1080/09500340008244048}{J. Mod. Opt. \textbf{47}, 355–376 (2000)}}.\label{06}
\item \textit{K. Zyczkowski, P. Horodecki, A. Sanpera, and M. Lewenstein,
``Volume of the set of separable states," \href{https://doi.org/10.1103/PhysRevA.58.883}{Phys. Rev. A \textbf{58}, 883 (1998)}}.\label{07}
\item \textit{K. Zyczkowski, ``Volume of the set of separable states. II," \href{https://doi.org/10.1103/PhysRevA.60.3496}{Phys. Rev. A \textbf{60}, 3496 (1999)}}.\label{08}
\item \textit{J. Lee, M. Kim, Y. Park, and S. Lee, ``Partial teleportation of entanglement in a noisy environment," \href{https://doi.org/10.1080/09500340008235138}{J. Mod. Opt. \textbf{47}, 2151–2164 (2000)}}.\label{09}
\item \textit{G. Vidal and R. F. Werner, ``Computable measure of entanglement," \href{https://doi.org/10.1103/PhysRevA.65.032314}{Phys. Rev. A \textbf{65}, 032314 (2002)}}.\label{10}
\item \textit{J. Eisert, ``Entanglement in quantum information theory," \href{https://arxiv.org/abs/quant-ph/0610253}{quant-ph/0610253 (2006)}}.\label{11}
\item \textit{W. K. Wootters, ``Entanglement of formation of an arbitrary state of two qubits," \href{https://doi.org/10.1103/PhysRevLett.80.2245}{Phys. Rev. Lett. \textbf{80}, 2245 (1998)}}.  \label{12}
\item \textit{J. Eisert and M. B. Plenio, ``A comparison of entanglement measures," \href{https://doi.org/10.1080/09500349908231260
}{J. Mod. Opt. \textbf{46}, 145–154 (1999)}}. \label{13}
\item \textit{P. G. Kwiat, E. Waks, A. G. White, I. Appelbaum, and P. H. Eberhard, ``Ultrabright source of polarization-entangled photons," \href{https://doi.org/10.1103/PhysRevA.60.R773}{Phys. Rev. A \textbf{60}, R773 (1999)}}. \label{14}
\item \textit{R. Rangarajan, M. Goggin, and P. Kwiat, ``Optimizing type-I
polarization-entangled photons," \href{https://doi.org/10.1364/OE.17.018920}{Opt. Express \textbf{17}, 18920–18933 (2009)}}.  \label{15}
\item \textit{D. F. V. James, P. G. Kwiat, W. J. Munro, and A. G. White, ``Measurement of qubits," \href{https://doi.org/10.1103/PhysRevA.64.052312}{Phys. Rev. A \textbf{64}, 052312 (2001)}}.\label{16}
\item \textit{C. Eltschka, G. Tóth, and J. Siewert, ``Partial transposition as a direct link between concurrence and negativity," \href{https://doi.org/10.1103/PhysRevA.91.032327}{Phys. Rev. A \textbf{91}, 032327 (2015)}}.\label{17}
\item \textit{W. K. Wootters, ``Entanglement of formation and concurrence," \href{}{Quantum Information \& Computation \textbf{1}, 27–44 (2001)}}.\label{18}
\item \textit{Y. Maleki and B. Ahansaz, ``Quantum correlations in qutrit-like superposition of spin coherent states," \href{https://doi.org/10.1088/1612-202X/ab12e5}{Laser Phys. Lett. \textbf{16}, 075205 (2019)}}. \label{19}
\item \textit{Y. Maleki and A. M. Zheltikov, ``Linear entropy of multiqutrit nonorthogonal states," \href{ttps://doi.org/10.1364/OE.27.008291}{Opt. express \textbf{27}, 8291–8307 (2019)}}.\label{20}
\item\textit{ D. Ghosh, T. Jennewein, and U. Sinha, ``Entanglement certification and quantification in spatial-bin photonic qutrits," \href{https://arxiv.org/abs/1909.01367}{arXiv:1909.01367 (2019)}}. \label{21}
\item \textit{D. M. Greenberger, M. A. Horne, A. Shimony, and A. Zeilinger, ``Bell’s theorem without inequalities," \href{https://doi.org/10.1119/1.16243}{Am. J. Phys. \textbf{58}, 1131–1143 (1990)}}. \label{22}
\item \textit{W. Dür, G. Vidal, and J. I. Cirac, ``Three qubits can be entangled in two inequivalent ways," \href{https://doi.org/10.1103/PhysRevA.62.062314}{Phys. Rev. A \textbf{62}, 062314 (2000)}}. \label{23}
\item \textit{H. J. Briegel and R. Raussendorf, ``Persistent entanglement in arrays of interacting particles," \href{https://doi.org/10.1103/PhysRevLett.86.910}{Phys. Rev. Lett. \textbf{86}, 910 (2001)}}. \label{24}
\item \textit{V. Coffman, J. Kundu, and W. K. Wootters, ``Distributed entanglement," \href{https://doi.org/10.1103/PhysRevA.61.052306}{Phys. Rev. A \textbf{61}, 052306 (2000)}}. \label{25}
\item \textit{A. Miyake, ``Classification of multipartite entangled states by multidimensional determinants," \href{https://doi.org/10.1103/PhysRevA.67.012108}{Phys. Rev. A \textbf{67}, 012108 (2003)}}. \label{26}
\item \textit{D. A. Meyer and N. R. Wallach, ``Global entanglement in multiparticle systems," \href{https://doi.org/10.1063/1.1497700}{J. Math. Phys. \textbf{43}, 4273–4278 (2002)}}. \label{27}
\item \textit{G. K. Brennen, ``An observable measure of entanglement for pure states of multi-qubit systems," \href{}{Quantum Information \& Computation \textbf{3}, 619–626 (2003)}}. \label{28}
\item \textit{Y. Maleki, F. Khashami, and Y. Mousavi, ``Entanglement of three spin states in the context of su (2) coherent states," \href{https://doi.org/I 10.1007/s10773-014-2215-5}{Int. J. Theor. Phys. \textbf{54}, 210–218 (2015)}}. \label{29}
\item \textit{Y. Maleki and A. Maleki, ``Entangled multimode spin coherent states of trapped ions," \href{https://doi.org/10.1364/JOSAB.35.001211}{J. Opt. Soc. Am. B \textbf{35}, 1211–1217 (2018)}}. \label{30}
\item \textit{G. Vidal and R. Tarrach, ``Robustness of entanglement," \href{https://doi.org/10.1103/PhysRevA.59.141}{Phys. Rev. A \textbf{59}, 141 (1999)}}. \label{31}
\item \textit{R. A. Horn and C. R. Johnson, Matrix Analysis (Cambridge University Press, 2012), \href{http://www.cambridge.org/9780521548236}{chap. 5, pp. 321}}. \label{32}
\item \textit{R. Jozsa, ``Fidelity for mixed quantum states," \href{https://doi.org/10.1080/09500349414552171}{J. Mod. Opt. \textbf{41}, 2315–2323 (1994)}}. \label{33}
\item \textit{D. Bures, ``An extension of kakutani’s theorem on infinite product measures to the tensor product of semifinite w*-algebras," \href{https://www.jstor.org/stable/1995012}{Trans. Am. Math. Soc. \textbf{135}, 199–212 (1969)}}. \label{34}
\item \textit{K. Zyczkowski and I. Bengtsson, ``Relativity of pure states entanglement," \href{https://doi.org/10.1006/aphy.2001.6201}{Ann. Phys. \textbf{295}, 115–135 (2002)}}.\label{35}
\item \textit{J. Anandan and Y. Aharonov, ``Geometry of quantum evolution," \href{https://doi.org/10.1103/PhysRevLett.65.1697}{Phys. Rev. Lett. \textbf{65}, 1697 (1990)}}. \label{36}
\item \textit{A. Gilchrist, N. K. Langford, and M. A. Nielsen, ``Distance measures to compare real and ideal quantum processes," \href{https://doi.org/10.1103/PhysRevA.71.062310}{Phys. Rev. A \textbf{71}, 062310 (2005)}.} \label{37}
\item \textit{N. Gisin, ``Hidden quantum nonlocality revealed by local filters," \href{https://doi.org/10.1016/S0375-9601(96)80001-6}{Phys. Lett. A \textbf{210}, 151–156 (1996)}.}  \label{38}
\item \textit{R. Horodecki, P. Horodecki, and M. Horodecki, ``Violating bell inequality by mixed spin-1/2 states: necessary and sufficient condition," \href{https://doi.org/10.1016/0375-9601(95)00214-N}{Phys. Lett. A \textbf{200}, 340–344 (1995)}}. \label{39}
\item \textit{B. Horst, K. Bartkiewicz, and A. Miranowicz, ``Two-qubit mixed states more entangled than pure states: Comparison of the relative
entropy of entanglement for a given nonlocality," \href{https://doi.org/10.1103/PhysRevA.87.042108}{Phys. Rev. A \textbf{87}, 042108 (2013)}.} \label{40}
\item \textit{S. Zohren and R. D. Gill, ``Maximal violation of the collins-gisinlinden-massar-popescu inequality for infinite dimensional states," \href{https://doi.org/10.1103/PhysRevLett.100.120406}{Phys. Rev. Lett. \textbf{100}, 120406 (2008)}.} \label{41}
\item \textit{C. Bernhard, B. Bessire, A. Montina, M. Pfaffhauser, A. Stefanov, and S. Wolf, ``Non-locality of experimental qutrit pairs," \href{https:/doi.org/10.1088/1751-8113/47/42/424013}{J. Phys. A \textbf{47}, 424013 (2014)}. } \label{42}
\item \textit{A. Acín, R. Gill, and N. Gisin, ``Optimal bell tests do not require maximally entangled states," \href{https://doi.org/10.1103/PhysRevLett.95.210402}{Phys. Rev. Lett. \textbf{95}, 210402 (2005)}. } \label{43}
\item \textit{N. Brunner, N. Gisin, and V. Scarani, ``Entanglement and nonlocality are different resources," \href{https://doi.org/10.1088/1367-2630/7/1/088}{New J. Phys. \textbf{7}, 88 (2005)}.} \label{44}
\item \textit{C. Eltschka and J. Siewert, ``Negativity as an estimator of entanglement dimension," \href{https://doi.org/10.1103/PhysRevLett.111.100503}{Phys. Rev. Lett. \textbf{111}, 100503 (2013)}}.  \label{45}
\end{enumerate}
\end{document}